\documentclass[doublecol]{epl2} 
\usepackage{amssymb}
\usepackage{graphicx}
\usepackage{caption}
\usepackage{amssymb}
\usepackage{subfig}
\usepackage{amsmath,amsthm}

\title{%Randomly pinned particles from intermediate temperature configurations: the birth of a new glass transition.
A general approach to systems with randomly pinned particles: unfolding and clarifying the Random Pinning Glass Transition} %clarifying explaining
\shorttitle{Origin of the Random Pinning Glass Transition} %Insert here a short version of the title if it exceeds 70 characters

\author{Chiara Cammarota\inst{1,2}}
\shortauthor{Cammarota}

\institute{                    
  \inst{1} Institut Physique Th\'eorique (IPhT) CEA Saclay, and CNRS URA 2306 - 91191 Gif Sur Yvette, France\\
  \inst{2} Universit\'e Pierre et Marie Curie - bo\^ite 121, 4 Place Jussieu, 75252 Paris c\'edex 05, France}
\abstract{
Pinning a fraction of particles %(or spins) 
from an equilibrium configuration in supercooled liquids %(or spins in suitable spin-glass systems) 
has been recently proposed as a way %procedure 
to induce a new kind of glass transition, %for the remaining free particles, 
the Random Pinning Glass Transition (RPGT). %, at temperature higher than $T_K$, where the putative thermodynamic glass transition takes place. 
The RPGT has been predicted to share some features of standard thermodynamic glass transitions %at $T_K$, if any, 
and usual first order ones. 
Thanks to its special nature, the approach and the study of the RPGT appears to be a fairly reachable task compared to the daunting problem of inspecting standard glass transitions. 
In this Letter we generalize the pinning particle procedure. 
We study %analyzing the case of 
a mean-field system where the pinned configuration is extracted from the equilibrium distribution at temperature $T'$ and the thermodynamics of the non pinned particles is observed at a lower temperature $T$. A more complicated physics emerges from this generalization eventually clarifying the origin and the peculiar characteristics of the RPGT.
}

\begin{document}

\maketitle

\section{Introduction}
Experimental observations% from more than sixty years ago
~\cite{Kauzmann, AnPoSh94, Cav_rev, BeBi_rev} on super-cooled liquids and plastic materials show a quite universal growth of relaxation time, which spans many order of magnitudes when temperature decreases in a rather restricted temperature window.
Soon enough the relaxation time is so large that it is not possible to experimentally equilibrate these systems. %at equilibrium, 
This equilibrium/non-equilibrium crossover occurs at a temperature $T_g$, the temperature of the so-called glass transition.\\
Despite a long lasting debate on static and dynamic approaches to the problem of glass transition~\cite{Gil_rev, ChaGar_rev}, the nature and the origin of the dynamical slowing down in super-cooled liquids remain an open problem.
According to one of the possible (static) explanations, %of this phenomenon %the huge growth of the relaxation time when temperature decreases 
the glass transition is the result of a particularly severe critical slowing down in proximity of 
%relies on the critical slowing down of 
a new thermodynamic transition, {\it i.e.} the ideal glass transition~\cite{KTW,LubWol_rev}, at a temperature $T_K<T_g$. The corresponding critical properties are characterized by 
%and characterized by new critical properties: notably 
an exponential relation between correlation time and cooperative length-scales. 
In agreement with this picture, it has been observed in numerical simulations~\cite{BBCGV,SauTar10,ProKar11,BeKo_PS,HocRei12,ChChTa12} that typical static length-scales cover only few inter-particle distances when relaxation time exceeds the observation time scale. In other words the putative universal behavior does not clearly emerge %is far from appearing 
in the temperature region where it is possible to equilibrate the system and a proper analysis of the critical properties cannot be achieved. Moreover, the whole low-temperature phase is out of reach of any possible equilibrium observation.
As a consequence, the mere existence of the glass transition and its position on the temperature axis, at a zero or finite temperature $T_K$, are matter of debate.
All these issues hamper in practice %the possibility of 
any stringent experimental test of this particular theory of the glass transition and, in general, the comparison between predictions coming from static and dynamic views. %is extremely difficult. 
%At the moment whether the transition is at finite temperature $T_K$ or at zero temperature and even whether it exists or not is not clear.
\\
Important results of the thermodynamic approach to glass transition have been recently obtained via the procedure of pinning a large fraction of particles in the system~\cite{SKBP,Kim1,Cavagnacavity1,BBCGV,CGVdyn,KoSaBe12,GrTrCa12,ParKar12,CB_RPGT}. This is a new tool% of statistical physics
~\cite{BouBir} specifically designed to study the properties of disordered systems, where the lack of a simple order parameter is a major issue. %of thermodynamic approaches. 
The idea consists in pinning a fraction of degrees of freedom from an equilibrium configuration at temperature $T$ and observing the thermodynamic of the remaining free particles.
This procedure has been successfully used to determine the existence of static cooperative length-scales in numerical models of molecular liquids~\cite{BBCGV, BeKo_PS}. %, when the pinned particles are chosen according to different geometries. 
Moreover, recently it has been suggested~\cite{CB_RPGT,CB_RPGTlong,CB_RPGTdyn} that for a particular realization of this procedure, {\it i.e.} when the $cN$ pinned particles are randomly chosen among the $N$ particles of a system, a genuine thermodynamic transition can be induced in three dimensions~\cite{Kim1, BeKo_PS} at temperature $T_K(c)>T_K$.
The interest in this second result is manifold.
First, this set of predictions can be only obtained within a specific thermodynamic theory of the glass transition called Random First Order Transition (RFOT) theory~\cite{KTW ,LubWol_rev}. Within other static theories and dynamic approaches invariantly is envisioned~\cite{CB_RPGT} and obtained~\cite{JacBer12} only the occurrence of a simple crossover instead of a transition. Hence, testing this prediction represents a new way to disclaim among different theories of glass transition.
Second, the nature of the new thermodynamic transition, called Random Pinning Glass Transition (RPGT), is not exactly the same as the usual ideal glass transition at $T_K$. 
In particular the RPGT shares features of both the ideal glass transition and standard first order transitions. It shows an exponential divergence of relaxation and correlation time in the liquid phase, but also a non-cooperative microscopic correlation time in the amorphous phase. 
There is no latent heat at the transition but, accordingly to Mean-Field computations, the amorphous low-temperature phase is constituted by a single amorphous minimum without any remnant of the more exotic one-step replica symmetry breaking ($1$RSB) phase that is conjectured to take place below the usual glass transition.
These features distinguish the RPGT as the first bona-fide glass transition that can be approached from both the liquid and the glass phase with substantial benefits for the inspection task.
Third, the now possible discovery of RPGT for a range of finite densities $c$ of pinned particles can eventually give information on the existence of the ideal glass transition for the free system and quantitative insights on its temperature position (either at a finite $T_K$ or at zero temperature).\\
The purpose of this Letter is to shed light on the origin of the RPTG and of its particular features that could ultimately enable the inspection of an ideal glass transition in real systems.
To this aim, we propose a generalization of the pinning particle procedure.
We analyze the case of a system where the pinned particles belong to an equilibrium configuration at temperature $T'$ and the thermodynamics of the remaining free particles is studied for temperatures $T<T'$.
In particular we will focus on the case $T'=T/\alpha $ with $0\le\alpha\le1$\footnote{The imposed condition on $T'$  
is such that the interesting case $T'=T$ where the RPGT appears is recovered for $\alpha=1$.}.
%easily obtain the phase diagram and the RPGT line in the interesting case $T'\rightarrow T^+$.
This generalization also provides new important predictions aimed at
%that can be directly tested in experiments and numerical simulations.
clarifying recent results~\cite{ParKar12} or promoting new tests in experiments and numerical simulations.\\
In practice, we present the results of replica computations in the spherical $p$-spin model, the ubiquitous spin model which provides a reliable mean-field description of supercooled liquids in proximity of the glass transition~\cite{Cavspin_rev,BoBi_rev,KTpspin,CriSom}.
In this case a fraction $c$ of spins is blocked in the positions that they have in an equilibrium configuration at temperature $T'$. The thermodynamics of the remaining spins is studied at $T<T'$.
Before presenting the results of this computation, we briefly review the outcomes earlier obtained in the same model using other methods to bias the thermodynamics and induce thermodynamic transitions.

\section{The $p$-spin model and early bias-induced thermodynamics transitions}
The spherical $p$-spin model~\cite{KTpspin} is a lattice spin model defined by the following Hamiltonian:
\begin{equation}
H=-\sum_{(i,j,\dots k)}J_{ij\dots k}s_is_j\dots s_k
\label{Hamilton}
\end{equation}
where the sum is performed over all the possible groups of $p$ spins in a system of $N$ spins, $J_{ij\dots k}$ are i.i.d. Gaussian random variables with zero mean and variance $\sigma_J=p!/2N^{p-1}$, and the spin $s_i$ are continuous variables bounded by the following spherical constraint: $N=\sum_{i=1}^Ns_i^2$.
At a temperature $T_K$, a singularity in the equilibrium free-energy identifies the ideal glass transition between a paramagnetic phase and the so-called $1$RSB low-temperature phase. This phase is characterized by a non trivial bimodal distribution of the order parameter $Q$, the similarity between two equilibrium configurations of the system.
The other characteristic temperature in the model is the dynamic transition temperature $T_d$. It marks the presence of a dynamical transition akin to the one found in Mode Coupling Theory. %; $T_d$ can be obtained as the highest temperature where a stationary point of the $1$RSB action with $q_1\neq q_0$ exists in the replica symmetric limit~\cite{CriSom,Cavspin_rev}.\\
\\
%The interest in studying the thermodynamics of a model with pinned spins is based on the particular nature of the RPGT and of its trivial low temperature phase, different from the $1$RSB phase.
As pointed out in~\cite{CB_RPGT}, the emergence of 
%such features 
the particular nature of the RPGT and of its trivial low temperature phase
is directly due to the special choice of the configuration used to pin the spins (an equilibrium configuration from the same temperature $T$) and to the pinning protocol itself.
Let's consider two other different procedures: the $\epsilon$-coupling approach by S. Franz and G. Parisi~\cite{FraPar97} and the procedure of pinning spins form a totally random configuration~\cite{CB_RPGTlong}.\\
In the first case, a glassy system at temperature $T$ is coupled with strength $\epsilon$ to a reference configuration, a typical equilibrium configuration of the same system at the same temperature. 
At variance with the pinning procedure, the coupling is introduced by adding an explicit term in the Hamiltonian that alters the equilibrium energy of the system.
For a small strength $\epsilon$, a purely first order transition~\cite{FraPar97} between a paramagnet (the equivalent of the liquid phase) and a trivial amorphous phase is observed in the system at temperature $T_s(\epsilon)>T_K$.\\
In the second case, a number $cN$ of spins among $N$ spins are frozen in a totally random configuration. A more complicated phase diagram emerges in this case~\cite{CB_RPGTlong}. At small $c$ a usual glass transition is induced at $T_K(c)>T_K$ towards a $1$RSB %low-temperature 
phase. At large $c$ a continuous transition between the $1$RSB phase and a trivial amorphous phase also takes place.\\
%According to the results presented in this Letter, 
The case of a system with particles pinned from an equilibrium configuration is strikingly different from the two previous examples. %: the continuous transition and the $1$RSB phase do not show up at all. Also the first-order transition and the original ideal glass transition do not appear, instead a new kind of glass transition, the RPGT, shows hybrid features from these two transitions.
A detailed analysis of the differences between the pinning procedure and the $\epsilon$-coupling approach have been recently presented in~\cite{CB_RPGTlong}.
The comparison between two pinning procedures, using equilibrium or random configurations, is even more interesting because it sheds light on the intrinsic hidden differences between low-temperature equilibrium configurations and random configurations.
The best way to approach this problem is to study the interpolating case of a system where particles are pinned from equilibrium configurations at temperature $T'=T/\alpha$, intermediate between the $T'=T$ ($\alpha=1$) case and the random choice $T'=\infty$ ($\alpha=0$). %, {\it i.e} the case of random configurations.
%We will see how the phase diagram changes, the $1$RSB low-temperature phase disappears and the RPGT is originated while the interpolating parameter $\alpha$ increases from zero to one.

%which separates the liquids phase from a single state amorphous phase at low temperature.
%replaces both the ideal glass transition and the first-order transition 
%disentangle entropy contribution and energy contribution
%To understand the origin of the RPGT and the hallmarks of the procedure able to induce this new kind of transition %we consider in this letter a generalization of the pinning particle procedure. 
%This prescription gives back in the two extremal cases, $\alpha=0$ and $\alpha=1$,  respectively the case where particles are pinned from a totally random configuration and the case where particles are pinned from equilibrium configuration at the same temperature T. Hence it is the suitable prescription to study how the $1$-SRB low-temperature phase evolves and disappears when $T'\rightarrow T^+$ and how the RPGT is originated.
%The new kind of glass transition presents at the same time the interesting features of a glass transition and the possibility of being approached from both phases.

\section{The $p$-spin model with a finite fraction of randomly pinned particles: the general case}

The free-energy of a $p$-spin model with a finite fraction of spins blocked from an equilibrium configuration is obtained as the result of a double average: first the average over the equilibrium configurations at temperature $T'$ of the pinned spins, then the average over different realizations of the couplings $\{J_{i,j,\dots,k}\}$.
The free-energy hence reads 
\begin{equation}
\mathcal{F}(c,T;T')=-\frac T N\overline{\sum_{\mathcal{C}_f}P^J({\mathcal{C}_f},T')\log Z^J_{\mathcal{C}_f}(c,T)} \ ,
\end{equation}
where $P^J({\mathcal{C}_f},T')=\exp(-\beta' H^J({\mathcal{C}_f}))/Z^J(T')$ is the equilibrium probability of the configuration $\mathcal{C}_f$ in the free system at temperature $T'=1/\beta'$; $Z^J_{\mathcal{C}_f}(c,T)$ is the partition function of the system at temperature $T$  where $cN$ spins at random are constrained to be in the same configuration as in the reference configuration $\mathcal{C}_f$, and $Z^J(T')=\sum_{\mathcal{C}}\exp(-\beta' H^J(\mathcal{C}))$ is the partition function for the free system at temperature $T'$. 
Introducing replicas in accordance with the procedure exposed in~\cite{CB_RPGTlong}, performing the double average and summing over all the configurations we obtain the following expression for the non-trivial part of the free-energy of the system:
\begin{eqnarray}
\mathcal{F}(c,T;T')=%-\frac{T}{Nn(m-1)}\lim_{\substack{m\rightarrow1\\n\rightarrow0\\N\rightarrow\infty}}\log\overline{(\sum_{\mathcal{C}_f}P^J({\mathcal{C}_f},T')(Z^J_{\mathcal{C}_f})^{m-1})^n}=
%\nonumber\\
\hspace{-0.2cm}
\lim_{
%\begin{array}{c}m\rightarrow1\\n\rightarrow0\\ N\rightarrow\infty\end{array}
\substack{m\rightarrow1\\n\rightarrow0\\N\rightarrow\infty}
} %\hspace{-0.2cm}
\frac {-T}{Nn(m-1)}\log\hspace{-0.15cm}
\int \hspace{-0.15cm}d\mathcal Q%_{a_{\alpha},b_{\beta}}
\exp[-NS(\mathcal Q)
%(Q_{a_{\alpha}b_{\beta}};c,T;T')
]\ ,
\end{eqnarray}
where 
\begin{eqnarray}
S(\mathcal Q)%(Q_{a_{\alpha}b_{\beta}};c,T;T')
=-\sum_{\alpha\beta}^{1,n}\left[\frac{\beta^2}{4}\hspace{-0.1cm}\sum_{a_{\alpha}\neq b_{\beta}}^{2,m}(c+(1-c)Q_{a_{\alpha}b_{\beta}})^p+\right.\nonumber\\
\left.+\frac{\beta\beta'}{2}\sum_{a_{\alpha}}^{2,m}(c+(1-c)Q_{a_{\alpha}1_{\beta}})^p\right]-\frac{(1-c)}{2}\log\det(\mathcal Q)%_{a_{\alpha}b_{\beta}})
\end{eqnarray}
and the $\mathcal Q$ matrix with elements $Q_{a_\alpha b_\beta}$ is an $nm$\hspace{0.05cm}x\hspace{0.05cm}$nm$ diagonal block matrix with $n$\hspace{0.05cm}x\hspace{0.05cm}$n$ blocks each one of $m$\hspace{0.05cm}x\hspace{0.05cm}$m$ elements. 
%\footnote{For the pinning construction with two temperatures, this is the simplest {\it ansatz} able to obtain a non-trivial low temperature phase.
%Actually to have access to the whole $c,T$ phase diagram, including the region where $T<\alpha T_K$, {\it i.e.} $T'<T_K$, and to fairly account for the reference equilibrium configuration $\mathcal{C}_f$ in this case, we would have to introduce a more complicated $1$RSB structure at the level of the block matrix, 
%which is by now simply diagonal.
%at the level of the configurations at temperature $T'$ meaning 
%} 
Each block has a special replica, say replica $1$,  equilibrated at temperature $T'$ and $m-1$ replicas at temperature $T$.
The off-diagonal blocks have all zero components.
The diagonal blocks have $1$ on the diagonal elements, the first row and column is characterized by the parameter $q$, and the remaining $(m-1)$\hspace{0.05cm}x\hspace{0.05cm}$(m-1)$ elements are shaped in a $1$RSB structure with parameter $q_1$, $q_0$, and $\mu$.
This structure of the overlap matrix brings into the problem all the relevant order parameters: $q$ is the overlap between the reference configuration $\mathcal C_f$ and the biased equilibrium configurations $\mathcal C$, $q_1 (q_0)$ is the overlap between two biased configurations belonging to the same (different) minimum and $\mu$ is the probability of finding biased configurations in different minima\footnote{This choice for $\mathcal Q$ properly restitutes a non-trivial $1$RSB phase for the biased system when $T'\geq T_K$. For $T'<T_K$ a more complicated $1$RSB structure instead of the diagonal block structure has to be considered for the $nm$\hspace{0.05cm}x\hspace{0.05cm}$nm$ matrix reflecting the $1$RSB structure of reference configurations.}.
%reference configuration $\mathcal C_f$ has to be introduced instead of the diagonal }.
According to this {\it ansatz} the action $S(Q_{ab};c,T;T')$ reads as follows:
\begin{eqnarray}
S(q,q_1,q_0,\mu%;c,T;T'
)=n(m-1)\left\{ \frac{\beta^2}{4}(1-\mu)(c+(1-c)q_1)^p\right. \nonumber\\ 
+\frac{\beta^2}{4}(\mu-(m-1))(c+(1-c)q_0)^p-\frac{\beta\beta'}{2}(c+(1-c)q)^p\nonumber\\
-\frac{1-c}{2}\left[\left(1-\frac{1}{\mu}\right)\log(1-q_1)+\frac{1}{\mu}\log(1-q_1+\mu(q_1-q_0))\right.\nonumber\\
\left.\left.+\frac{q_0-p^2}{1-q_1+\mu(q_1-q_0)}\right]\right\} \ .
\end{eqnarray}
Finally, the equilibrium free-energy and the physical values of the parameters $q,q_1,q_0$ and $\mu$ will be obtained taking the saddle point value of $S$, which amounts to solve the following set of equations:
\begin{subequations}
\label{ST}
\begin{align}
\label{STq}
\alpha\frac{\beta^2}{2}p(c+(1-c)q)^{p-1}=\frac{q}{1-q_1+\mu(q_1-q_0)}\\
\label{STq0}
\frac{\beta^2}{2}p(c+(1-c)q_0)^{p-1}=\frac{q_0-q^2}{(1-q_1+\mu(q_1-q_0))^2}\\
\frac{\beta^2}{2}p(c+(1-c)q_1)^{p-1}=\frac{1}{\mu}\left[\frac{1}{1-q_1}-\frac{1}{1-q_1+\mu(q_1-q_0)}\right.\nonumber\\
\label{STq1}
\left.+\mu\frac{q_0-q^2}{(1-q_1+\mu(q_1-q_0))^2}\right]\\
0=\frac{\beta^2}{2}\left[(c+(1-c)q_1)^p-(c+(1-c)q_0)^p\right]\nonumber\\
+\frac{(1-c)}{\mu}\left[\frac{1}{\mu}\log\left(\frac{1-q_1}{1-q_1+\mu(q_1-q_0)}\right)\right.\nonumber\\
\label{STm}
\left.+\frac{(q_1-q_0)}{1-q_1+\mu(q_1-q_0)}\left(1-\mu\frac{q_0-q^2}{1-q_1+\mu(q_1-q_0)}\right)\right] \ ,
\end{align}
\end{subequations}
where $\beta'$ has been replaced by $\alpha\beta$.
These equations have two different classes of solutions. The first one is the replica symmetric (RS) solution with $\mu=1$, and $q^{RS}_0=q_1^{RS}\neq q^{RS}$, meaning that the overlap between two randomly chosen biased configurations $\mathcal C$s is always $q_1$ while the overlap between one of them and the reference configuration $\mathcal C_f$ is $q$. %even if $T'\neq T$ means $q\neq q_1$,
%, which is the stable one at high temperature/small concentration $c$, 
The second one is the $1$RSB solution with $\mu^{1RSB}<1$, $q^{1RSB}_0<q^{1RSB}_1$, and $q^{1RSB}<q^{1RSB}_1$, implying an overlap $q$ between $\mathcal C$ and $\mathcal C_f$ and two possible overlap values between two $\mathcal C$s: $q_0$ with probability $\mu$ and $q_1$ with probability $1-\mu$. 
%, the stable one at low temperature/high concentration.
The appearance of the $1$RSB solution with $\mu=1$ defines the dynamic transition line $T_d(c)$~\cite{CrHoSo93,CugKur93}.
The glass transition is found at the temperature where the free-energies of the RS and the $1$RSB solutions become equal: $\mathcal{F}^{RS}=\mathcal{F}^{1RSB}$ or when equations~\eqref{ST} are satisfied by the $\mu=1$ condition. This defines a glass transition line $T_K(c)$ in the $c,T$ phase diagram.\\
Generalizing the procedure in~\cite{CriSom}, we also found a solution to equations \eqref{ST} in the limit $q_1-q_0\rightarrow0^+$: %which in this case corresponds also to the limit $q-q_0\rightarrow0^+$, 
%in this case we have
\begin{subequations}
\label{STcontinuous}
\begin{align}
\label{STcontq}
\alpha\frac{\beta^2}{2}p(c+(1-c)q)^{p-1}=\frac{q}{1-q_0}\\
\label{STcontq0}
\frac{\beta^2}{2}p(c+(1-c)q_0)^{p-1}=\frac{q_0-q^2}{(1-q_0)^2}\\
\label{STcontT}
\frac{\beta^2}{2}p(p-1)(1-c)(c+(1-c)q_0)^{p-2}=\frac{1}{(1-q_0)^2}\\
\label{STcontm}
m=\frac{(p-2)(1-c)(1-q_0)}{2(c+(1-c)q_0)} \ .
\end{align}
\end{subequations}
The set of equations \eqref{STcontinuous} defines the continuous transition line $T_c(c)$ in the physically meaningful range $0<m<1$ corresponding to $\frac{1}{2}\frac{p-2}{p-1}\leq q_0\leq1$. 
Actually, the continuous transition line spans %is physically meaningful only 
an even smaller region.
Through a careful examination, we found that the $1$RSB solution, that gives rise to the continuous transition, is metastable in certain regions of the phase diagram. %not everywhere on this line the free-energy has its lowest possible value hence the continuous transition line is physically meaningful only on an even more restricted range.
The reason is that
a third kind of thermodynamic transition occurs in the general case $0<\alpha<1$. 
It is a first-order transition line, taking place when increasing $c$, between a low-$q$ $1$RSB phase %at low-$c$ 
and a high-$q$ RS phase or, for small values of $\alpha$, between two $1$RSB phases with low and high $q$.\\
The three lines of thermodynamic transition arrange in a quite elaborate phase diagram for intermediate $\alpha$ and disappear or merge in the extreme cases $\alpha=0$ and $\alpha=1$.
In the following section we present a detailed analysis of the phase diagrams for two values of $\alpha=0.5, 0.9$ and we refer to~\cite{CB_RPGTlong} for the phase diagrams of the extreme cases $\alpha=0,1$.
The dynamic transition line $T_d(c;\alpha)$, indicating where the metastable $1$RSB solution with $\mu=1$ appears, will be also present in the phase diagrams. 

\section{The phase diagrams}
\begin{figure}
\subfloat[Phase diagram for $\alpha=0.5$]{\label{fig:alpha0.5}\includegraphics[width=.48\textwidth,angle=0]{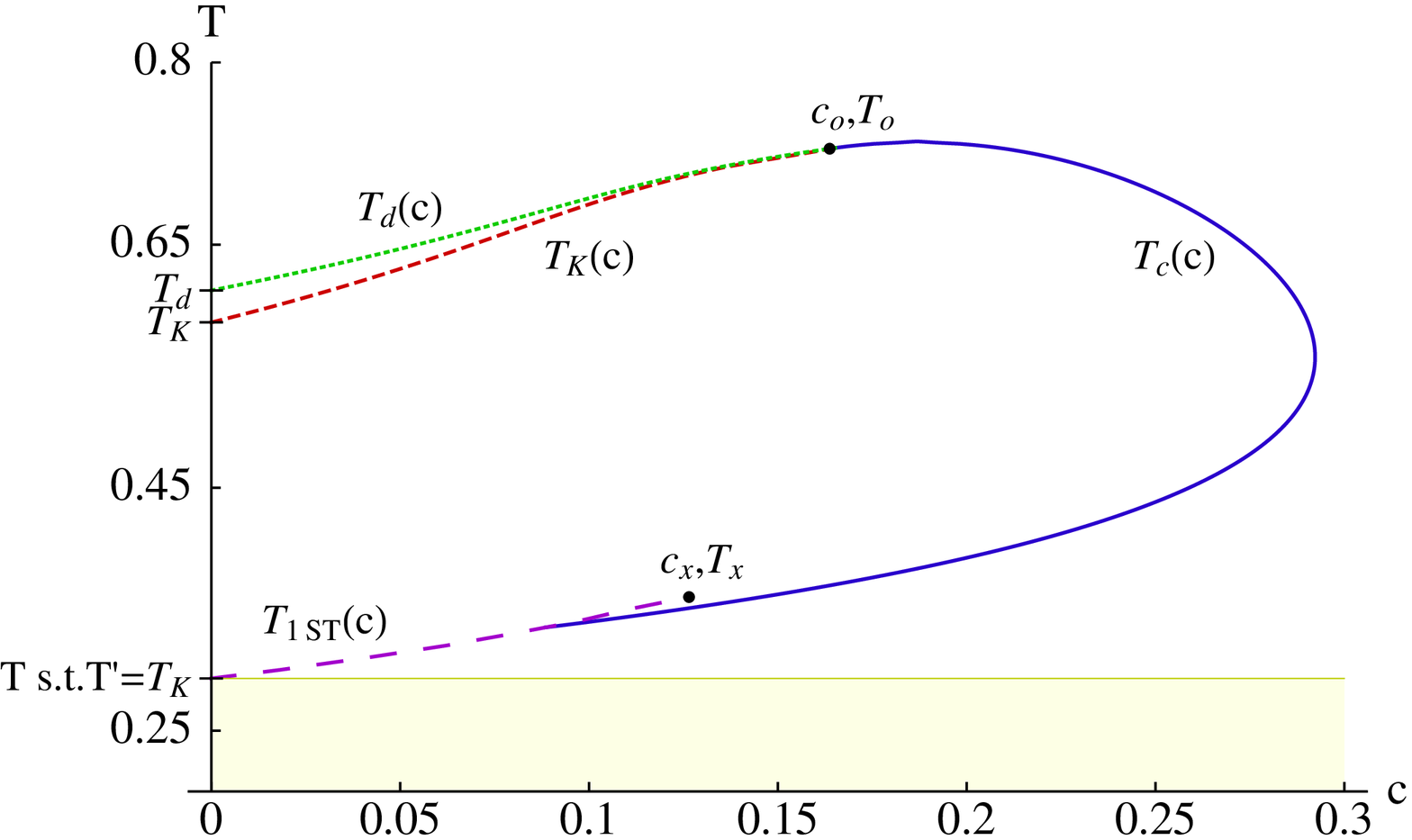}}\vspace{-0.5cm}\\
\subfloat[Phase diagram for $\alpha=0.9$]{\label{fig:alpha0.9}\includegraphics[width=.48\textwidth,angle=0]{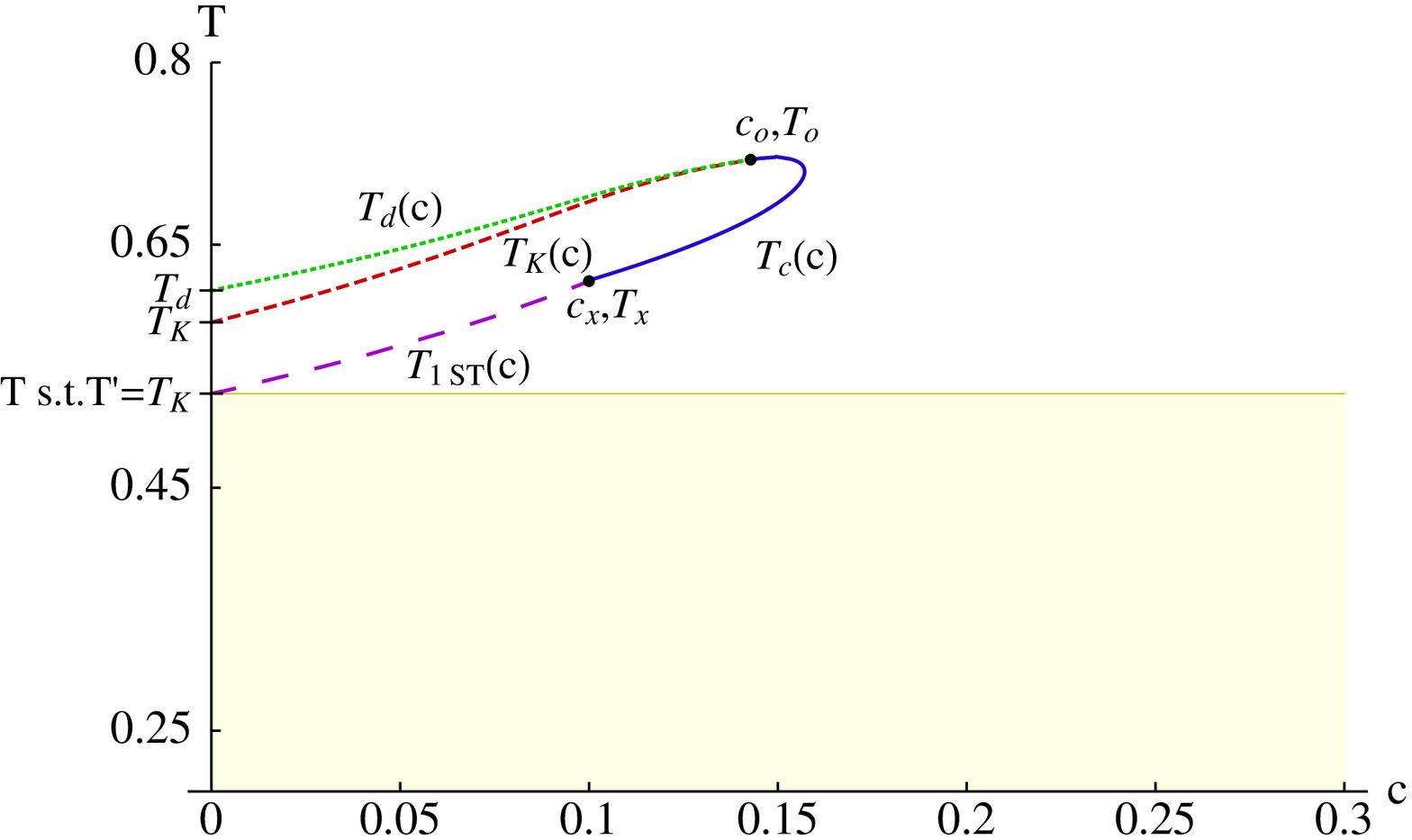}}
\caption{Comparison of two phase diagrams for different values of the parameter $\alpha\in(0,1)$. A simple first-order transition appears for small $\alpha$s, cutting the continuous transition line. %When $\alpha$ increases this originates the shrinking of the $1$RSB phase and the contraction of the continuous transition line. 
Eventually the first-order transition line and the discontinuous glass transition line superimpose giving rise to the RPGT line in the case $T'=T$.}
\label{fig:phasediagrams}
\vspace{-0.3cm}
\end{figure}

In this section we show how the phase diagram changes when $\alpha$ varies from $0$ to $1$.
We illustrate in particular the way the $1$RSB phase shrinks and the continuous transition line contracts to eventually give rise to the peculiar $\alpha=1$ phase diagram presented in~\cite{CB_RPGT,CB_RPGTlong}, where the replica symmetric phase is everywhere thermodynamically stable and a new kind of discontinuous glass transition emerges. \\
The phase diagram in the $\alpha=0$ case~\cite{CB_RPGTlong} is qualitatively similar to the one obtained for the $p$-spin model in a random field~\cite{CriSom,CrHoSo93}. In fact, a very strong local random field has the same effect as pinning a fraction of spins. %In fact, the randomly pinned spins play the role of a local random field applied to a mixed $p$-spin model with effective interactions among $p, p-1, \dots 2$ free spins.
As usual in this case, a $1$RSB phase covers the low temperature region of the phase diagram.
Referring to the parameters introduced here to describe the most general case, the $1$RSB phase is characterized by $q_1\geq q_0$, $q=0$, and $0<\mu\leq 1$.
This phase is closed off  by a discontinuous glass transition line $T_K(c)$, in every respect similar to the ideal glass transition at $T_K(c=0)$ where $\mu=1$ and $q_1> q_0$, and by a continuous glass transition line $T_c(c)$ where $\mu<1$ and $q_1-q_0\rightarrow0^+$.
At the two sides of this region two different RS phases lie: a paramagnetic phase (the analogous of the liquid phase for particle systems) with low $q_1=q_0$ in the leftmost part and a frozen amorphous phase selected by the pinned particles for high concentrations $c$ with high $q_1=q_0$. In both cases we have $q=0$.  \\
For a non-zero $\alpha$, {\it e.g.}  in the $\alpha=0.5$ case shown in Fig.\ref{fig:alpha0.5}, the $1$RSB phase is found to be shrunk in favor of the frozen amorphous RS phase as consequence of a folding of the $T_c(c)$ continuous transition line, while the $T_K$ and $T_d$ lines are slightly shifted.
%On top of this, a qualitatively different feature emerges due to a non-zero equilibrium value of $q$. % the presence of a non-zero $q$ solution. 
On top of this, another $1$RSB phase is found with $q_1>q\gtrsim q_0$. This solution is actually the stable one in the low-temperature, low-concentration part of the phase diagram being the extension of the usual $1$RSB phase with $q_1\neq0$ and $q=q_0=0$ of the unpinned system, we call it $1\text{RSB}_{\text{UP}}$ phase.
When $c$ increases this $1\text{RSB}_{\text{UP}}$ solution remains the stable one till a first order transition (the $T_{1ST}(c)$ line in the phase diagram) takes place in favor of either the high-overlap RS phase or a $1$RSB phase with $q_1>q_0> q$ akin to the $1$RSB phase in the $\alpha=0$ case. 
Note that the first-order transition line limit for $c=0$ is at a temperature $T$ such that %the corresponding $T'$ is 
$T'=T_K$\footnote{A $T'=T_K$ equilibrated $\mathcal C_f$ belongs to a well shaped metastable minimum among few others. An infinitesimal bias in this direction naturally selects it giving immediately rise to a RS phase.
%an infinitesimal concentration of spins of an equilibrium configuration at $T_K$ is enough to confine the thermodynamics of the system in a single well shaped metastable state among few others, hence equilibrium is immediately given by a RS phase.
}.\\
%Indeed, for $T'>T_K$ the $c=0$ equilibrium solution is the usual $1$RSB solution of the free system 
%In this case an infinitesimal bias $c=\epsilon$ is able to thermodynamically favor the RS frozen phase. ...
One main consequence of the presence of the new $1\text{RSB}_{\text{UP}}$ phase is that the continuous transition line defined by equations~\eqref{STcontinuous} is thermodynamically relevant only where the $1$RSB solution with $q_1>q_0>q$ is stable, {\it i.e.} on the right of the $T_{1ST}$ line. On the left, the solution with $q_1-q_0\rightarrow 0^+$ pertains to the metastable branch of the free-energy, hence it does not emerge in the equilibrium behavior of the system%and we did not trace it on the phase diagram
\footnote{Following the metastable $1$RSB solution with $q_1>q_0>q$, we observed that the $T_c$ line ends at $c=0$ on a temperature $T$ such that $T'=T_d$. %Indeed, %This has a clear explanation in terms of the free-energy landscape. F
%for $T'>T_d$ the reference configuration does not necessarily belong to a well shaped metastable state, hence the biased system %by this configuration 
%will come across a non-trivial $1$RSB phase before the RS frozen amorphous phase is reached. On the contrary, a biasing configuration equilibrated at $T'<T_d$ will immediately drive the system in a single-minimum solution, even if metastable.
}. Two critical points appear in this case: $c_o,T_o$ between the discontinuous and the continuous glass transition lines, and $c_x,T_x$, a second-order transition point at the end of the first-order transition line. 
\\
As the parameter $\alpha$ increases towards one, see the $\alpha=0.9$ case in Fig.\ref{fig:alpha0.9}, the phase diagram does not change qualitatively but it clearly shows the way the phase diagram in the $\alpha=1$ case will be re-obtained.
The $1$RSB region shrinks, the continuous glass transition line shortens, and the first-order transition approaches the discontinuous glass transition line.
The first-order transition line ends on the continuous glass transition. %and only a single $1$RSB phase remains. 
\\
In the $\alpha=1$ limiting case the two discontinuous transitions and the two critical points superimpose and the continuous line disappears\footnote{Also in this case we can follow the continuous transition line on the metastable branch. We obtain in this way that it superimposes to the dynamic transition line $T_d(c)$. Indeed for $\alpha=1$ this is the line where metastable states form in the free-energy landscape of the pinned system. Hence it corresponds to the line where the structure of the still metastable solution with high $q$ becomes RS.}.
As a result, the $1$RSB region vanishes and only one line of glass transitions remains in the phase diagram. This line represents a new kind of glass transition, which has been firstly presented in~\cite{CB_RPGT}. 
For every $\alpha<1$, the $T_K(c)$ line represents the usual glass transition between a paramagnetic phase, with a single low-overlap value between biased configurations, and a glass phase. This second phase is characterized by a double peak probability distribution function of the order parameter, where the weight of the high overlap peak continuously grows from $0$ as the transition is left behind. Only beyond the continuous or the first-order transition line a RS phase with a single value of the overlap between the biased configurations can be recovered.
On the contrary, when the region of the $1$RSB phase vanishes and the first-order transition line merges with the usual glass transition line, the overlap distribution function is never double peaked and the resulting transition is associated to a genuine jump of the overlap from a single low value to a high one.
For $\alpha=1$ the $T_K(c)$ line becomes a line of strongly discontinuous glass transitions. %This becomes evident looking at %the typical order parameter of disordered system, the similarity (or overlap) 
%the overlap between two equilibrium configuration, $Q$.
%The probability distribution function shows in this case a low-overlap single peak in the paramagnetic phase on the left of the RPGT line and a high overlap single peak on the right side, in the frozen amorphous phase.
%On the contrary, usual (discontinuous) glass transitions are associated to a zero overlap peak in the liquid phase and a non-trivial double peak structure in the glass phase, which is actually a $1$RSB phase. In particular the high-overlap peak appears at a finite $Q$ but its weight continuously grows from $0$ as temperature decrease below the glass transition. Hence, to mark the difference with respect to usual glass transition, we label the RPGT as strongly discontinuous.
As a result the low-temperature/high-concentration phase %, a RS-kind phase, 
is constituted by a simple minimum selected by the reference configuration of pinned particles. It is already well formed when it becomes stable, as it usually happens in first order phase transitions.
On the other hand, the physics in the paramagnetic part of the phase diagram is the one that emerges in proximity of usual glass transition: it is characterized by an exponential number of metastable states and a well defined configurational entropy that decreases when temperature decreases and vanishes at the transition~\cite{CB_RPGTlong}. %Hence it is in every respect similar to the one which precedes usual glass transitions.
\\
Finally, we note that: first, for $T=T'$ we obtain the interruption in a critical point of the RPGT line 
already claimed in~\cite{CB_RPGT}. In our study the critical point is originated by the merging of $c_0,T_0$ and $c_x,T_x$. Second, every pair of points in the $T=T'$ phase diagram can be naturally connected by a continuous path that does not cross any thermodynamic singularity\cite{CB_RPGT} because the disconnected $1$RSB phase has been squeezed on the RPGT line.
%These features naturally appear in this Letter as the consequence of the peculiar superposition of a glass transition transition of usual kind and a first order transition in the $T=T'$ limit.

\section{Conclusion}

The study of a $p$-spin system at temperature $T$ where a fraction $c$ of spins are blocked in an equilibrium configuration at temperature $T'$ provides intermediate results between the case of random pinning of spins from an equilibrium configuration~\cite{CB_RPGT} and from a completely random configuration~\cite{CB_RPGTlong}.
A first order transition line appears in this general case. It always interrupts the continuous glass transition line already present in the completely random configuration case~\cite{CriSom,CB_RPGTlong} and eventually it merges, for $\alpha\rightarrow1$, with the usual glass transition line (between a RS and a $1$RSB phase) giving rise to the RPGT line when $T=T'$.\\
In this paper we imposed $T<T'$. The opposite choice, $T'<T$, or $\alpha>1$, gives rise to a much simpler phase diagram: a first order transition between the paramagnetic/liquid phase and the frozen amorphous phase intervenes before the glass transition has been attained, similarly to what happens in the $\epsilon$-coupling approach. The present RS {\it ansatz} for the reference configurations gives valuable results  %$c,T$ phase diagram extends %well before $T_K$, {\it i.e.} 
in the temperature $T$ range such that $T'>T_K$.
Despite the little practical relevance, it could be at least theoretically interesting to extend the results and the picture we have obtained in this Letter to the region $T'<T_K$. This can be done introducing a replica symmetry breaking scheme for the reference configurations used to pin particles in the system.
\\
As usual with the $p$-spin glass model, mean-field results can be extrapolate to obtain predictions (to be tested) for real liquids within the framework of a landscape-controlled picture of the glass transition, we devote the last part of this section to this issue. \\
In real systems the dynamic transition becomes a crossover, we expect the same for the dynamic transition line for every value of the parameter $\alpha$. On the contrary, the other thermodynamic transition lines should be present also in finite dimensions $D$ (at least for $D\ge3$), thought they will be characterized by completely different critical properties.
%We surmise the typical finite-dimensional mechanisms for the slowing down of the dynamics in the liquid phase when glass transitions (the usual one or the RPGT) are approached. 
Both the usual glass transition line and the RPGT line are induced in the system by the vanishing of the configurational entropy~\cite{CB_RPGT,CB_RPGTlong}. Hence, according to the RFOT~\cite{KTW,LubWol_rev}, activated events involving a larger and larger number of particles should lead to an exponential growth of the relaxation time and of the equilibrium correlation time. 
%We also expect that the first-order transition is present in finite dimensions (at least in three dimensions). 
%At variance with what happens near the glass transition, i
At variance, in proximity of the first-order transition line, the equilibrium correlation time has to remain almost constant. %in the amorphous frozen phase. 
Only the time scale required to relax from non-equilibrium configurations diverges and the phenomenology associated to metastability and nucleation events emerges.
%Finally real systems should show the existence of the continuos glass transition line. It 
Finally, the continuous glass transition line could be associated to a critical power-law divergence of relaxation and correlation time when it is approached from the frozen amorphous phase in the rightmost part of the phase diagram.
In the intermediate region a non-trivial glass phase should be found where equilibrium configurations can only be obtained as a result of a formidable optimization task for medium size systems and are impossible to be determined in the large size limit. Correlation time and relaxation time would be invariantly infinite in this region as it is theoretically expected for the glass phase of non-pinned glass-formers below $T_K$.
In experiments in $2D$ or $3D$ colloidal systems, thanks to the technology of the optical tweezers, or in molecular liquids (even if it is yet unclear how to find a way to pin particle), the simplest observables to probe are correlation and relaxation time, and average overlap, hence this set of predictions can be directly tested.
Note that, in the $\alpha=1$ case the expectation is that only for $D\geq 3$ the RPGT is a true transition and in $D=2$ it becomes a cross-over due to the intrinsic disorder of the system. A particularly interesting question is 
what changes in the $\alpha<1$ phase diagram where also continuous transitions are present
%whether continuous transitions remains for $\alpha<1$
\footnote{In~\cite{FrPaVi94} it is argued that the lower critical dimension of continuous spin-glass transition is $D_L=2.5$. The continuous transition discussed in this Letter has a different nature, hence the question on the physics in $D=2$ systems remains open.}.\\
%In experiments in $2$ or $3$D colloidal systems, thanks to the technology of the optical tweezers, or in molecular liquids (even if it is quite a hard work to find a way to pin particle in that case), the simplest observables to probe are correlation and relaxation time. In the $\alpha=1$ case the divergence of correlation time is expected only on the RPGT line~\cite{CB_RPGT} (and only for $D\leq 3$), preceded by an exponential growth in the liquid phase and an almost constant correlation time in the frozen amorphous phase. Approaching the transition in this phase only the time scale required to relax from non-equilibrium configurations diverges and the phenomenology associated to metastability and nucleation events emerges.
%Analogous dynamical phenomenology as in the $\alpha=1$ case is expected for $\alpha<1$ in the liquid phase near the glass transition line and in the frozen amorphous phase when the first-order transition line is approached. 
Even if equilibrium is not always attainable in the thermodynamic limit, %in some particular cases, 
finite-size scaling studies could give access to equilibrium results in every region of the phase diagram. In particular, among the possible static observables, it is worth to look at the probability distribution function of the overlap between equilibrium configurations.
This has already been done in a recent numerical work~\cite{ParKar12} that we can naturally ascribe to the $\alpha\sim0$ case and in a numerical work-in-progress~\cite{BeKo_PQ} on the $\alpha=1$ case. 
Supporting evidences of both the continuous glass transition and the RPGT emerge from these works.
%The continuous and the discontinuous glass transition find their first verification
Further numerical studies on the intermediate $\alpha$ cases would be suitable to utterly validate the predictive skill of this thermodynamic approach to the glass transition.\vspace{-0.6cm}\\

%In this Letter, we proposed a generalization of the pinning particle procedure.
%A particular construction that lead to a glass transition of novel kind, called Random Pinning Glass transition 

%\begin{figure}
%\includegraphics[width=0.45\textwidth]{alpha0.5.eps}\\
%\includegraphics[width=0.45\textwidth]{alpha0.9.eps}
%\caption{blabla}
%\label{figures}
%\end{figure}

\acknowledgments
I thank G. Biroli, A. Crisanti, S. Franz, L. Leuzzi, E. Marinari, G. Parisi, and G. Tarjus for interesting discussions. Special thanks to A. Cavagna and G. Biroli to whom I am indebted for their precious support and attentive guidance during PhD and first years of post-doc. I acknowledge support from the ERC grant NPRGGLASS. 
%Special thanks to Andrea Cavagna and Giulio Biroli for their unique 
% efforts in my formation 
%help in the course of my first steps as a researcher.

\bibliography{bib}	

\begin{thebibliography}{10}
\expandafter\ifx\csname url\endcsname\relax\def\url#1{\texttt{#1}}\fi

\bibitem{Kauzmann}
\Name{Kauzmann W.} \REVIEW{Chemical Reviews}{43}{1948}{219}.

\bibitem{AnPoSh94}
\Name{Angell C.~A., Poole P.~H. \and Shao J.} \REVIEW{Nuovo Cimento
  D}{16}{1994}{993}.

\bibitem{Cav_rev}
\Name{Cavagna A.} \REVIEW{Physics Reports}{476}{2009}{51}.

\bibitem{BeBi_rev}
\Name{Berthier L. \and Biroli G.} \REVIEW{Reviews of Modern
  Physics}{83}{2011}{587Ð645}.

\bibitem{Gil_rev}
\Name{Tarjus G.} \Book{Dynamical Heterogeneities in Glasses, Colloids, and
  Granular Media} Vol. Chapter 2: An overview of the theories of the glass
  transition (Oxford University Press) 2012.

\bibitem{ChaGar_rev}
\Name{Chandler D. \and Garrahan J.~P.} \REVIEW{Annual Review of Physical
  Chemistry}{61}{2010}{191}.

\bibitem{KTW}
\Name{Kirkpatrick T.~R., Thirumalai D. \and Wolynes P.~G.} \REVIEW{Physical
  Review A}{40}{1989}{1045Ð1054}.

\bibitem{LubWol_rev}
\Name{Lubchenko V. \and Wolynes P.~G.} \REVIEW{Annual Review of Physical
  Chemistry}{58}{2007}{235}.

\bibitem{BBCGV}
\Name{Biroli G., Bouchaud J.-P., Cavagna A., Grigera T.~S. \and Verrocchio P.}
  \REVIEW{Nature Physics}{4}{2008}{771}.

\bibitem{SauTar10}
\Name{Sausset F. \and Tarjus G.} \REVIEW{Physical Review
  Letters}{104}{2012}{065701}.

\bibitem{ProKar11}
\Name{Karmakar S. \and Procaccia I.} \REVIEW{arXiv:1105.4053}{}{2011}{5pgs}.

\bibitem{BeKo_PS}
\Name{Berthier L. \and Kob W.} \REVIEW{Physical Review E}{85}{2012}{011102}.

\bibitem{HocRei12}
\Name{Hocky G.~M., Markland T.~E. \and Reichman D.~R.} \REVIEW{Physical Review
  Letters}{108}{2012}{225506}.

\bibitem{ChChTa12}
\Name{Charbonneau B., Charbonneau P. \and Tarjus G.} \REVIEW{Physical Review
  Letters}{108}{2012}{035701}.

\bibitem{SKBP}
\Name{Scheidler P., Kob W., Binder K. \and Parisi G.} \REVIEW{Philosophical
  Magazine A}{82}{2002}{283}.

\bibitem{Kim1}
\Name{Kim K.} \REVIEW{Europhys. Lett.}{61}{2003}{790}.

\bibitem{Cavagnacavity1}
\Name{Cavagna A., Grigera T.~S. \and Verrocchio P.} \REVIEW{Physical Review
  Letters}{105}{2010}{055703}.

\bibitem{CGVdyn}
\Name{Cavagna A., Grigera T.~S. \and Verrocchio P.} \REVIEW{J. Chem.
  Phys.}{136}{2011}{204502}.

\bibitem{KoSaBe12}
\Name{Kob W., Rold\'an-Vargas S. \and Berthier L.} \REVIEW{Nature
  Physics}{8}{2012}{164}.

\bibitem{GrTrCa12}
\Name{Gradenigo G., Trozzo R., Cavagna A., Grigera T.~S. \and Verrocchio P.}
  \REVIEW{arXiv:1209.5954}{}{2012}{}.

\bibitem{ParKar12}
\Name{Karmakar S. \and Parisi G.} \REVIEW{arXiv:1208.3181}{}{2012}{}.

\bibitem{CB_RPGT}
\Name{Cammarota C. \and Biroli G.} \REVIEW{Proceedings of the National Academy
  of Sciences}{109}{2012}{8850}.

\bibitem{BouBir}
\Name{Bouchaud J.-P. \and Biroli G.} \REVIEW{Journal of Chemical
  Physics}{121}{2004}{7347}.

\bibitem{CB_RPGTlong}
\Name{Cammarota C. \and Biroli G.} \REVIEW{arXiv:1210.8399}{}{2012}{28pgs}.

\bibitem{CB_RPGTdyn}
\Name{Cammarota C. \and Biroli G.} \REVIEW{Europhysics
  Letters}{98}{2012}{16011}.

\bibitem{JacBer12}
\Name{Jack R.~L. \and Berthier L.} \REVIEW{Phys. Rev. E}{85}{2012}{021120}.

\bibitem{Cavspin_rev}
\Name{Castellani T. \and Cavagna A.} \REVIEW{J. Stat. Mech.}{}{2005}{P05012}.

\bibitem{BoBi_rev}
\Name{Biroli G. \and Bouchaud J.-P.} \Book{Structural glasses and supercooled
  liquids: theory, experiment and applications} (Wiley) 2012.

\bibitem{KTpspin}
\Name{Kirkpatrick T.~R. \and Thirumalai D.} \REVIEW{Physical Review
  B}{36}{1987}{5388Ð5397}.

\bibitem{CriSom}
\Name{Crisanti A. \and Sommers H.-J.} \REVIEW{Z. Phys. B-Condensed
  Matter}{87}{1992}{341}.

\bibitem{FraPar97}
\Name{Franz S. \and Parisi G.} \REVIEW{Physical Review
  Letters}{79}{1997}{2486}.

\bibitem{CrHoSo93}
\Name{Crisanti A., Horner H. \and Sommers H.-J.} \REVIEW{Z. Phys. B-Condensed
  Matter}{92}{1993}{257}.

\bibitem{CugKur93}
\Name{Cugliandolo L.~F. \and Kurchan J.} \REVIEW{Physical Review
  Letters}{71}{1993}{173}.

\bibitem{FrPaVi94}
\Name{Franz S., Parisi G. \and Virasoro M.~A.} \REVIEW{Journal de Physique I
  France}{4}{1994}{1657}.

\bibitem{BeKo_PQ}
\Name{Berthier L. \and Kob W.} \REVIEW{private communication}{}{2012}{}.

\end{thebibliography}
\bibliographystyle{eplbib.bst}

%\begin{thebibliography}{0}

%\bibitem{b.a}
%  \Name{Author F., Author S. \and Author T.}
%  \REVIEW{Some Rev. A}{69}{1969}{9691}.

%\bibitem{b.b}
%  \Name{Author F. \and Author S.}
%  \Book{Some Book of Interest}
%  \Editor{A. Editor}
%  \Vol{9}
%  \Publ{Publishing house, City}
%  \Year{1939}
%  \Page{666}.

%\bibitem{b.c}
%  \Editor{Editor A.}
%  \Book{Some Book of Interest}
%  \Vol{9}
%  \Publ{Publishing house, City}
%  \Year{1939}
%  \Section{A}.

%\end{thebibliography}

\end{document}